\documentclass[twocolumn,showpacs,preprintnumbers,amsmath,amssymb]{revtex4}
\usepackage{graphicx}
\usepackage{dcolumn}
\usepackage{bm}
\usepackage{amssymb}
\usepackage{amsmath}
\begin{document}
\title{Current dependent fluctuations in a Bi$_2$Sr$_2$CuO$_{6+\delta}$ thin film}

\author{I. Sfar (*), Z.Z. Li, F. Bouquet, H. Raffy, L. Fruchter}%
\affiliation{Laboratoire de Physique des Solides, C.N.R.S.
Universit\'{e} Paris-Sud, 91405 Orsay cedex, France (*) also at L.P.M.C., D\'{e}partement de Physique,
Facult\'{e} des Sciences de Tunis, campus universitaire 1060
Tunis, Tunisia.}
\date{Received: date / Revised version: date}
%
\begin{abstract}{The current dependence of the excess conductivity is measured up to $\simeq 3\;T_c$ for a Bi$_2$Sr$_2$CuO$_{6+\delta}$ thin film, as a function of doping. It is found to be anomalously sensitive to the transport current and to behave as a universal function of $T/T_c$ in the whole doping range. We discuss these results in the perspective of a granular superconductor with a gapless-like behavior.}
\end{abstract}


\maketitle
\section{Introduction}

Superconducting fluctuations are strongly reduced when the out of equilibrium superfluid velocity due to a transport current reaches a critical value, $\Delta/p_F$, similar to the depairing velocity obtained in the superconducting regime.
Using the time dependent Ginzburg Landau theory, Schmidt and Hurault computed the associated critical electrical field in the case of Gaussian fluctuations\cite{schmidt1969,hurault1969,tsuzuki1970}.
Improvements for layered materials\cite{varlamov1992,mishonov2002} or taking into account the critical regime close to the superconducting transition temperature were obtained later\cite{dorsey1991,puica2003}. Clear
experimental evidence for the validity of the theories in the
Gaussian regime were brought in the case of bulk conventional
superconductors\cite{kajimura1970,thomas1971,kajimura1971a,kajimura1971b}.
However, similar studies on high-$T_c$ superconductors are rare,
principally due to the experimental difficulty to reach higher
critical electrical fields in these
materials\cite{soret1993,gorlova1995,fruchter2004}. In Ref.
\cite{fruchter2004}, an anomalously large sensitivity of the
superconducting fluctuations to the transport current was pointed
out for Bi$_2$Sr$_2$CuO$_{6+\delta}$ (Bi-2201), which results in an
apparent characteristic electrical field several orders of
magnitude lower than would be expected from a simple estimate for this material. The origin of this
discrepancy is still unclear\cite{fruchter2004}. Among possible
explanations, the existence of microscopic disorder, at a
length scale smaller than the one of the superconducting
fluctuations, was proposed. In this contribution, we
explore further the effect of the transport current for a Bi-2201
thin film, extending the non-linearity measurements up to $T \simeq
3\,T_c$ and from overdoped to underdoped superconducting states. We discuss the results in the
perspective of a granular material.

\section{Experiments}
\begin{figure}
\resizebox{0.7\columnwidth}{!}{\includegraphics{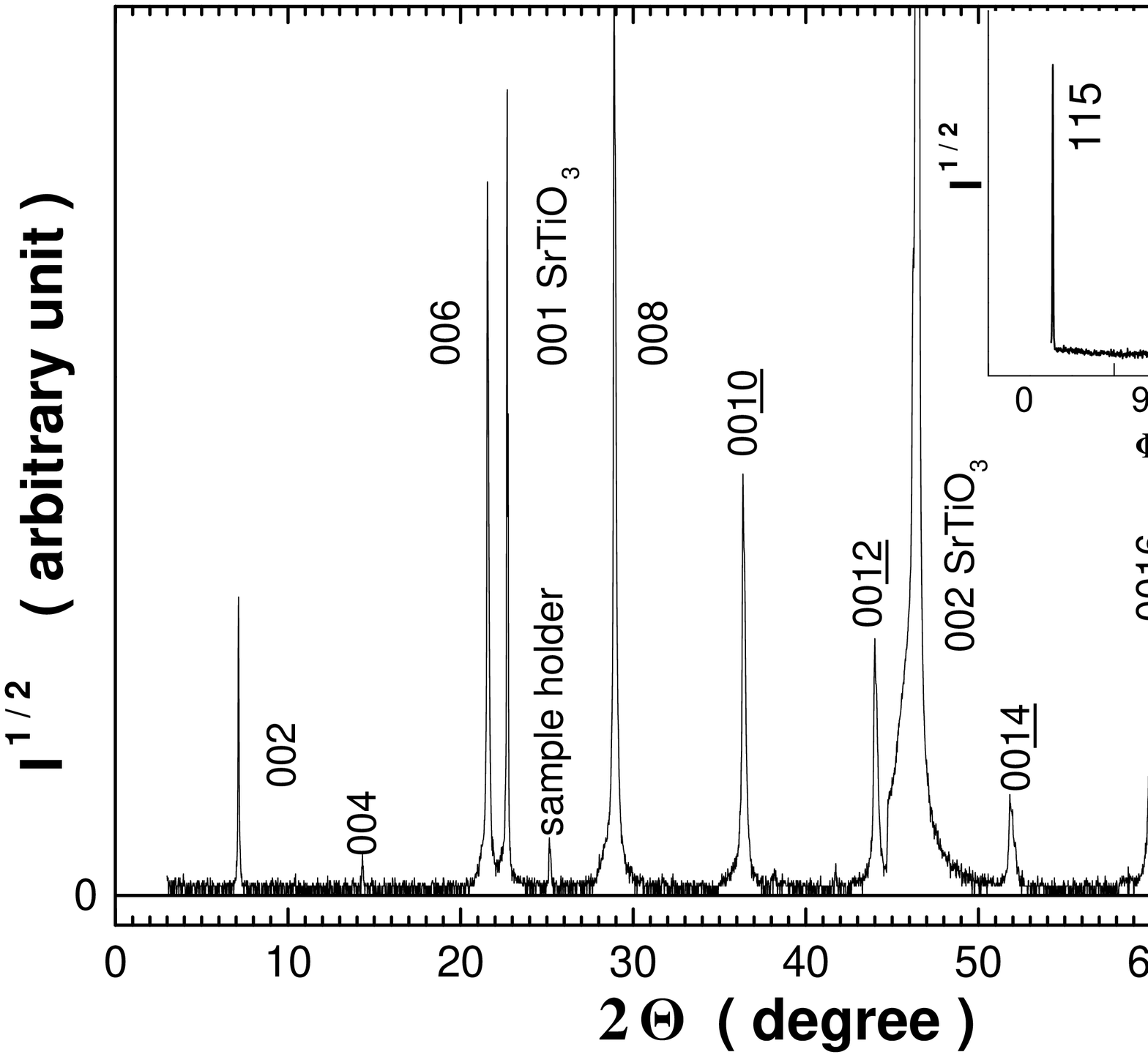}}
\caption{X-ray diffraction spectra for the Bi-2201 film. $\Theta-2\Theta$ scan. Inset : $\Phi$ scan  of the (115) reflection of the film.}\label{xray}
\end{figure}

A single crystal, c-axis oriented, Bi-2201 thin film was grown
epitaxially (Fig. \ref{xray}, inset) on a heated SrTiO$_3$ substrate, by reactive
\textit{rf} sputtering with an oxygen rich plasma
(Ref.~\cite{li1993} and Refs therein). It consists of grains with
a c-axis perpendicular to the film, with sharp twin boundaries at
the atomic level and no phase shift between them, due to the
orientation imposed by the SrTiO$_3$ substrate. X-ray diffraction analysis
also allowed us to check the absence of parasitic phases, to the
accuracy of the diffraction spectra, i.e. about 3\% (Fig. \ref{xray}). Resistive
measurements, which are sensitive to the presence of superconducting
intergrowth phases appearing as a kink in $dR/dT$ curves, did not
show any of these. This is expected in the case of Bi-2201 for
which no such phases are observed. After deposition of
2700~\AA{} thick material, Au contacts were sputtered onto the sample,
which was patterned in the four contact transport geometry,
with a current carrying strip of width and length 100~$\mu$m and
130~$\mu$m respectively. The orientation of the strip was such that the
current flew along the CuO bond direction. Doping was varied
by changing the oxygen content of the film. The film was annealed for one
hour at 270 $^\circ$C under the appropriate oxygen pressure. The sample resistance was monitored, allowing us to characterize in situ the variation of the sample doping level. The resistance rapidly stabilized and stayed constant, thus insuring the thermal equilibrium of the oxygen content. The sample was then rapidly
quenched to the ambient temperature. Doing so, we
obtained different doping levels for the \textit{same} sample,
with a 10\% -- 90\% resistive transition about 2 K wide. The
maximum superconducting temperature, measured at the mid-point of
the transition, was $T_c =$ 19.9 K.

The sample resistance at vanishing current density was measured
using a lock-in detection with a current of 10~$\mu$A. Larger current
resistance measurements, below $I=30$~mA (current density
$J=1.2\;10^5\;$A cm$^{-2}$), were performed using the pulse-probe
technique described in Ref. \cite{fruchter2004}. The current was
fed into the sample during a 10~$\mu$s pulse and, 1~$\mu$s
later, a probe pulse with a lower current $I_0 = 2$~mA and
negligible Joule heating was used to measure the sample resistance
and its temperature, while the thermal relaxation since the main
pulse is negligible. The repetition rate was $10^{-4}$. The temperature increase determined in this
way was less than 0.3~K for the largest current value.
The measured non linearity due to the electronics of the
experimental setup was 0.3\% for the higher current. Such a
value is not negligible as compared to the non linearity of the
sample conductivity. As a consequence, a correction was made for
the sample resistance, using exactly the same procedure for all
data shown below. It consisted in a normalization of the data, so
that the corrected sample resistance was independent of the
current in the range 110~K -- 120~K. We note that this procedure
may eliminate the non linearity which may be present at
temperature higher than this range. However, the non linearity
uncovered by this procedure being strongly increasing with
decreasing temperature, this validates \textit{a posteriori} the
low temperature data.
The heating of the sample is a major problem in these experiments. In our case, the transport
current needed to suppress the fluctuations is relatively small, as compared to other high-Tc samples \cite{fruchter2004,kunchur1995,lang2005}, so that the sample temperature rise is also small. As a consequence, we are clearly not in the situation where the superconducting transition appears as shifted due to heating by several Kelvin \cite{kunchur1995,lang2005}. Also, the fast temperature decrease at the end of the main pulse - which cannot be measured by our technique and is set by the thermal resistance at the film/substrate interface and by the dissipated power per unit area (70 W/cm$^2$ in our case, as compared to $10^4$ W/cm$^2$ for the thinner film in ref. \cite{lang2005})- is reduced to a few hundredth of Kelvin and could be neglected, while higher power would either require a specific temperature correction\cite{kunchur1995} or pulses shorter than the film relaxation time. The following results validate \textit{a posteriori} the procedure used to evaluate the sample temperature, as severe uncorrected heating effects would make the apparent non linear field effect independent of the doping level - which is not observed here - and as a less disordered sample consistently exhibits a reduced non linearity.

\label{results}

\section{Results and discussion}

\begin{figure}
\resizebox{0.7\columnwidth}{!}{%
  \includegraphics{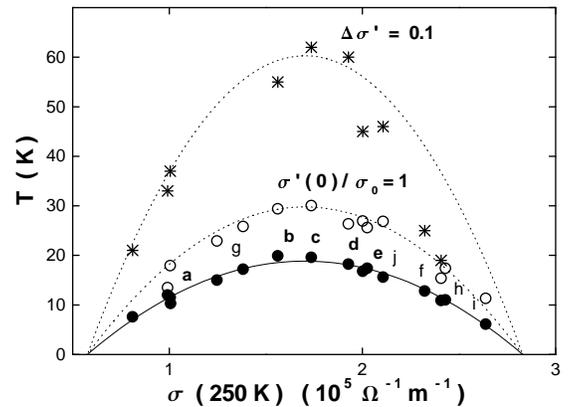}
}
\caption{Full circles: resistive transition temperature;
hollow circles --- respectively stars : temperature at which the excess
conductivity --- respectively the excess conductivity non linearity
--- meets a given criterion. Line, parabolic fit for the transition
temperature. Dotted lines are homothetic to the full one.}\label{Tcr}
\end{figure}

\begin{figure}
\resizebox{0.7\columnwidth}{!}{%
  \includegraphics{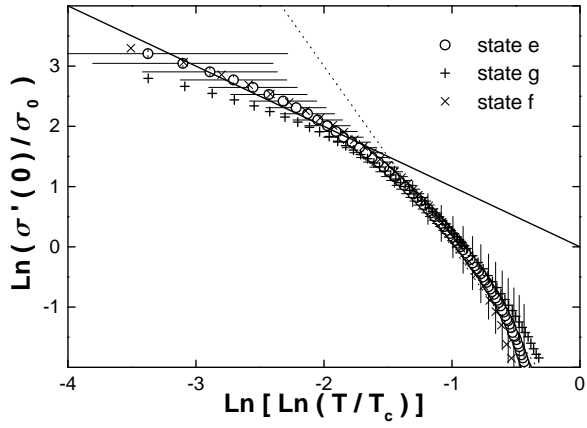}
}
\caption{\label{fluctuations}Excess conductivity in the vanishing current limit vs
reduced temperature, in the 2D universal conductance unit, using
$s= 12.3$~\AA. States are labelled according to Fig. \ref{Tcr}
notations (states e and f : overdoped $T_c$ = 17.4~K and 12.8~K;
state g : underdoped, $T_c$ = 17.2~K). The full line is the
Aslamazov-Larkin result\cite{aslamazov1968} and the dotted one is
the high temperature asymptotics in Ref. \cite{varlamov1999}.
Horizontal error bars originate from the finite transition width
and the vertical ones from the uncertainty on the normal state
conductivity.}
\end{figure}

\begin{figure}
\resizebox{0.5\textwidth}{!}{%
  \includegraphics{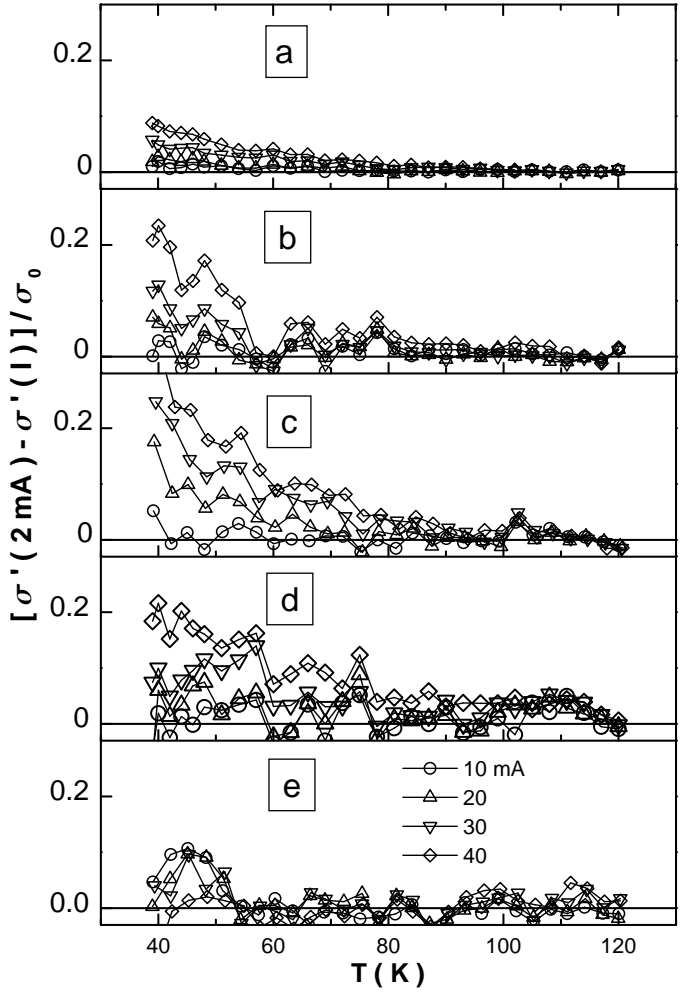}
}
\caption{Excess conductivity non linearity for various currents $I$. Doping states as shown
in Fig. \ref{Tcr} (from underdoped, top, to overdoped, bottom).}\label{nonlinearite}
\end{figure}

\begin{figure}
\resizebox{0.7\columnwidth}{!}{%
  \includegraphics{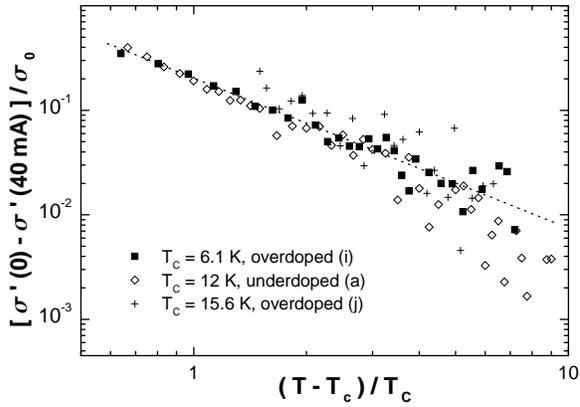}
}\caption{Universality of the-non linear excess conductivity.
The line slope is $-1.4$.} \label{universality}
\end{figure}

\begin{figure}
\resizebox{0.7\columnwidth}{!}{%
  \includegraphics{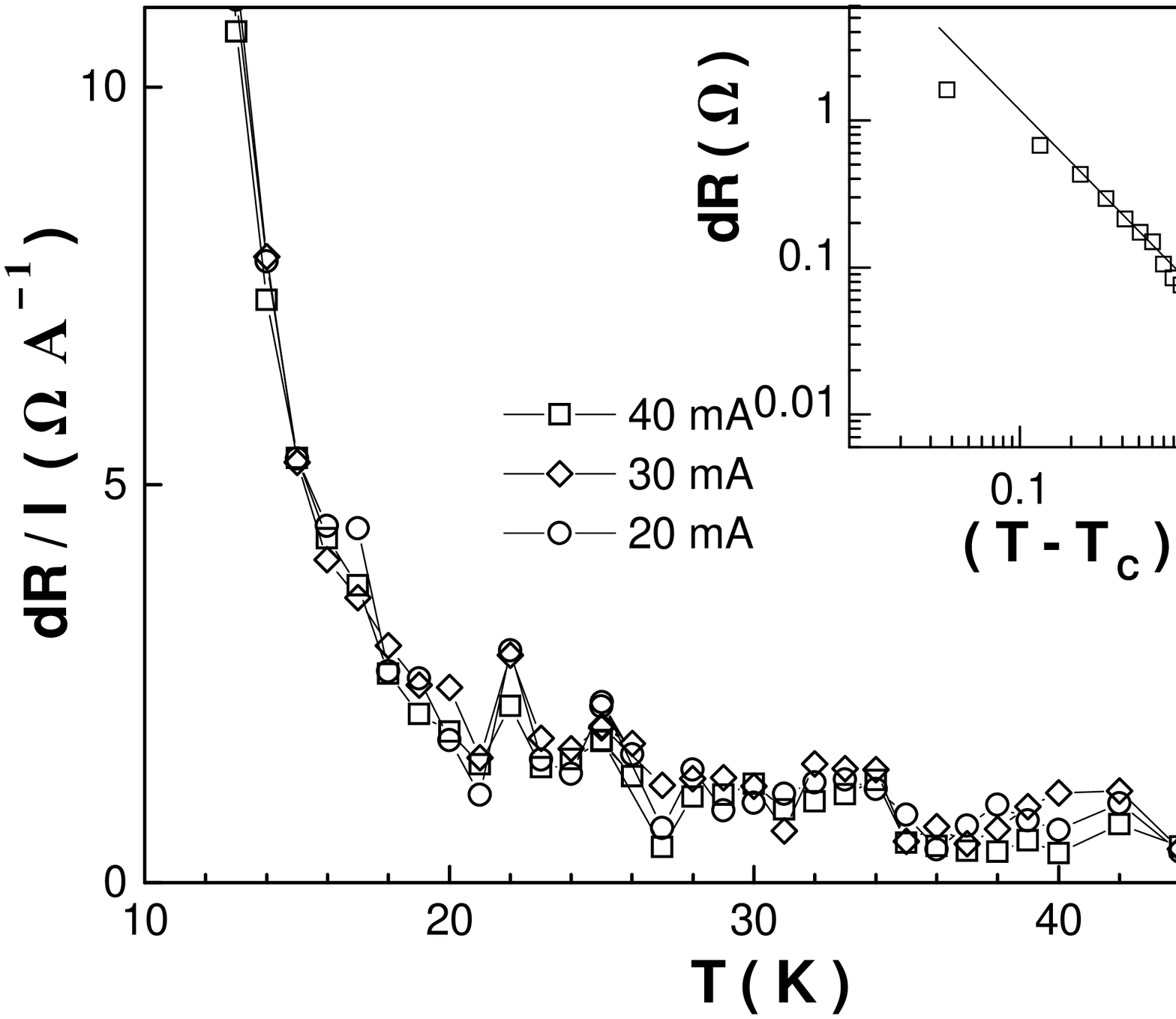}
}
\caption{Sample resistance variation, $dR = R(I)-R($2 mA$)$,
scaling as $I\;(T-T_c)^{-1.2}$ (overdoped state h, $T_c = 10.6$
K). Inset: I = 40 mA, line slope is $-1.2$.} \label{dR}
\end{figure}

The superconducting transition temperature for various doping
levels is shown in Fig. \ref{Tcr}, as a function of the sample
conductivity at 250~K, $\sigma(250$~K$)$. As one has roughly
$\sigma \propto p \propto \delta$, where $p$ is the CuO$_2$
plane hole concentration, the data may be fitted using the parabolic
empirical law \cite{takagi1989,presland1991}: $T_c/T_c^{op} = 1-
C(p-p^{op})^2$, where $T_c^{op}$ and $p^{op}$ are the
superconducting transition temperature and hole concentration for
optimal $T_c$. The excess conductivity in the limit of vanishing
current density, $\sigma '(0)$, with respect to the normal-state
conductivity as obtained from a fit of the normal-state
resistance above $2 \; T_c$ to a power law, is shown in
Fig. \ref{fluctuations} (the excess conductivity is defined as $\sigma'=\sigma-\sigma_{normal}$). As well known for this procedure, the
uncertainty on the excess conductivity is essentially due, on the
low temperature side, to the finite transition width and, on the
high temperature one, to the uncertainty for the normal-state
conductivity. Within these limitations, the excess conductivity
for both underdoped and overdoped states is found roughly
universal (i.e. dependent on the reduced temperature $T/T_c$
only) and well described by the two dimensional Aslamazov-Larkin
theory for Gaussian fluctuations (A-L)\cite{aslamazov1968} and the
high-temperature extension in
Ref. \cite{reggiani1991,varlamov1999}, \textit{with no fitting parameter}.
Equivalently, the temperature at which the excess conductivity
meets some criterion should be proportional to the sample
transition temperature only. Taking $\sigma '$ = $\sigma_0$, where
$\sigma_0 = e^2 / 16 \hbar s$ is the universal fluctuation
conductance, with $s$ the superconducting CuO$_2$ plane
separation, one may check in Fig. \ref{Tcr} the universal
character of the fluctuations on the whole range of doping. The
large current measurements (Fig. \ref{nonlinearite}), allow ones to
further uncover some weaker fluctuations at higher temperature.
Evaluating the temperature at which the excess conductivity for
the higher current is reduced by about $\Delta \sigma'\equiv\sigma'(2 mA)-\sigma'$(40 mA) =
$10^{-1}\,\sigma_0$, we find that, within the experimental
uncertainty, the fluctuations uncovered by the current are again
universal (Fig. \ref{Tcr}). We note, in particular, that the
current dependent fluctuations do not exhibit any enhancement on
the underdoped side of the phase diagram with respect to the
overdoped one (Fig. \ref{Tcr}. There is a slight asymmetry, which may be due to
the fact that our criterion is obtained at constant
\textit{current}, while a constant \textit{electrical field}
criterion would shift the points on the overdoped side -- with a
lower resistivity -- to higher temperatures). The universality of
the current dependent part of the excess fluctuations holds up to
a reduced temperature as high as $T \simeq 3\;T_c$, as can be seen
in Fig. \ref{universality}. From the sample resistance variation, $R(I)-R(I \rightarrow 0)$, as shown in Fig. \ref{dR}, one may evaluate the characteristic electrical field for the excess
conductivity non linearity, within the Gaussian theory. For $E \lesssim 0.3\,E_c(T)$, one has
$\sigma'(E)/\sigma'(0) \simeq 1-0.6\;(E/E_c)$
\cite{schmidt1969,hurault1969,tsuzuki1970}. Taking $\rho(T,E) \simeq \rho_n$ the normal-state resistivity (one has $(\rho-\rho_n)/\rho_n < 10^{-1}$ for
$\epsilon=(T-T_c)/T_c > 0.2$), we obtain:
\begin{equation}
[\rho(E)-\rho(0)]/\rho_n^2 \simeq \sigma'(0) - \sigma'(E) \simeq
0.6 \;\sigma_0 \epsilon^{-1} (E/E_c)
 \label{eq1}
\end{equation}

\begin{figure}
\resizebox{0.7\columnwidth}{!}{%
  \includegraphics{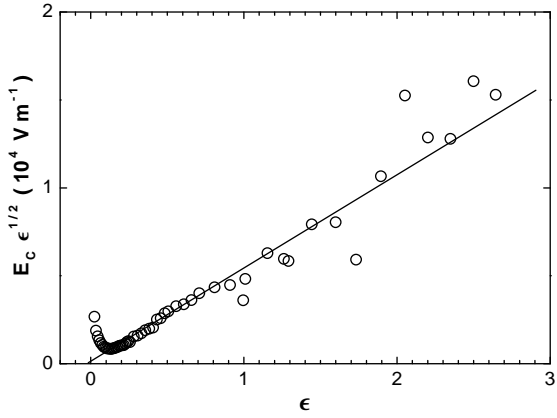}
}
\caption{\label{Ec}Apparent characteristic electrical field from
Eq. \ref{eq1} in a $\epsilon^{1/2}$ representation (state c, optimally doped). Line is a guide to the eye.}
\end{figure}

Then, for weak variations of the sample resistivity with current
(so that $E \propto I$), one expects $R(I)-R(0) \propto
\epsilon^{-5/2} I$, using $E_c\propto \epsilon^{3/2}$. As shown
in Fig. \ref{dR}, we do observe such a linear dependence with the current.
However, the temperature dependence is clearly weaker, being
close to $\epsilon^{-\alpha}$ with $\alpha = 1.2 - 1.5$
(Figs \ref{universality}, \ref{dR}). The apparent critical
electrical field obtained from the resistance variation is then
found increasing with temperature as $\sim \epsilon^{\alpha-1}$ (a
dependence clearly weaker than the theoretical one,
$\epsilon^{\,3/2}$) with a typical value $E_c(40$ K$) \simeq
3\;10^3$ V m$^{-1}$ (Fig. \ref{Ec}). This is well below the
expected value $(16 \sqrt{3} k_B T_c/ \pi e \xi_0)\epsilon^{3/2}
\simeq 3\,10^6 \epsilon^{\,3/2}$~V~m$^{-1}$ for the two dimensional
Gaussian case \cite{schmidt1969,hurault1969,tsuzuki1970}, using $\xi_0 = 60$~\AA{} \cite{triscone1997}, which
extends the results obtained in the interval $\epsilon \lesssim
0.5$ in Ref. \cite{fruchter2004} to higher temperatures.

\begin{figure}
\resizebox{0.7\columnwidth}{!}{%
\includegraphics{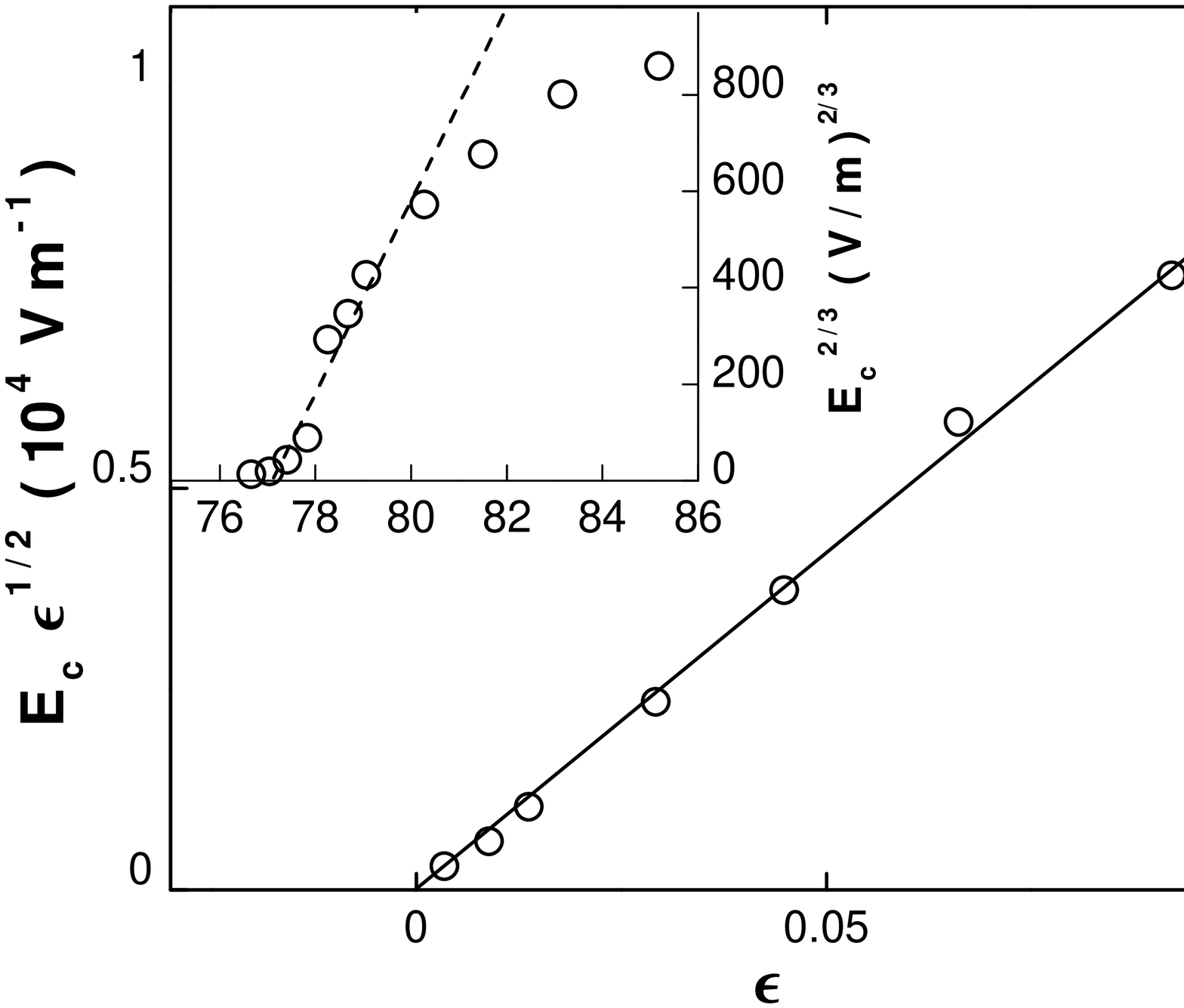}
} \caption{\label{gorlova}Characteristic electrical field for the
onset of non linearity of a Bi$_2$Sr$_2$CaCu$_2$O$_{8+\delta}$
single crystal from \cite{gorlova1995} in a $\epsilon^{1/2}$
representation with $T_c$ = 78 K. The inset is the same data as presented in \cite{gorlova1995}
and the dotted line accounts for the theoretical $\epsilon^{3/2}$
behavior.}
\end{figure}

There is little other experimental data on high-$T_c$
superconductors to compare with our measurements. Non linearity
was measured for YBa$_2$Cu$_3$O$_{7+\delta}$ in Ref.
\cite{soret1993}. However, the temperature range ($\epsilon <
0.02$) was too narrow to allow for a comparison. In
Ref. \cite{gorlova1995}, characteristic electrical field measurements
were reported for Bi$_2$Sr$_2$CaCu$_2$O$_{8+\delta}$ (Bi-2212)
with $T_c \simeq$ 78 K in a larger temperature range ($\epsilon <
0.1$). The electrical field was found somewhat smaller than
expected and a coherence length value as high as $\xi$ = 100 --
200~\AA{}  must be used to account for the data. Furthermore, the data
is clearly better described as $E_c \propto \epsilon^{1/2}$,
rather than by the conventional behavior $E_c \propto
\epsilon^{3/2}$ proposed in Ref. \cite{gorlova1995} (Fig. \ref{gorlova}).
Then, although the excess conductivity in Bi-2212 exhibits a smaller electrical field
dependence than in Bi-2201, it is still larger than expected (in
agreement with the results in Ref. \cite{fruchter2004}) and the
temperature dependence of the effect is found similar to the one
described in this paper for Bi-2201.

We shall now discuss these results in the framework of a granularity. As a granular superconductor may exhibit arbitrarily small critical current density, one might also expect
that a transport current can reduce superconducting fluctuations more easily than in the bulk. Ideally, such a material consists of identical superconducting grains surrounded by an insulator or
a normal metal, so that the grains are coupled through junctions.
In the present case, granularity should not be understood as the
presence of well defined grains with sharp boundaries, as can be
found in conventional granular materials, but as the presence of
inhomogeneities or `islands', such as the ones observed in
Refs. \cite{cren2000,pan2001,howald2001,lang2002}. There has already
been several proposals to account for anomalous properties of
some high-$T_c$ superconductors --- positive curvature of $H_{c2}(T)$,
Meissner and Nernst effects above $T_c$ --- using
granularity\cite{spivak1995,geshkenbein1998}. The following
considerations all pertain to the case of an s-wave
superconductor, whereas Bi-2201 materials is likely a d-wave one.
It is known that tunnel junctions from such materials may greatly
differ from the s-wave case. However, in the
large temperature limit, the phase space around the gap nodes
scales with the temperature, so that the d-wave junction should
essentially behave as conventional ones\cite{bruder1995}.

To begin with the vanishing-current resistivity measurements, one
may wonder whether the observation of standard universal
fluctuations (Fig. \ref{fluctuations}) is compatible with a disordered material. As noticed
in \cite{varlamov1999}, the universality of the fluctuations for
a two dimensional system is robust against all sorts of
perturbations, such as impurities or localization and this is
likely true also for a two dimensional granular superconductor
(see the discussion by Harris for the case of the two-dimensional
XY model with disorder \cite{harris1974} and
Ref. \cite{dalidovich2000} for a modelization with an array of
resistively shunted junction with moderate dissipation --- actually
isomorphic to the first case, as well as the fluctuation
conductivity obtained in Ref. \cite{deutscher1974} in the 3D case).
As an illustration, granular NbN films transport properties
were investigated in Ref. \cite{wolf1981}. The conductivity was found
to behave as $\sigma\propto(T-T_{cj})^{-3.7}$ over one decade,
where $T_{cj}$ is the temperature for phase ordering of the
superconducting network. This is in complete disagreement with
the A-L prediction $\sigma\propto (T-T_c)^{-1}$. However, a close
examination of the data in Ref. \cite{wolf1981} shows that there is
indeed a temperature regime, $(T-T_{cj})/T_{cj}< 0.1$, where the
A-L result is observed. The high-temperature power law may then
correspond to the asymptotics in
Refs. \cite{reggiani1991,varlamov1999}. Thus, we may conclude that
standard conductance fluctuations can be preserved in a two
dimensional granular superconductor (as long as the inhomogeneity
is weak enough, so that the superconducting transition is not
dominated by percolation\cite{char1988}), and the present low-current measurements do not rule out granularity in our case.

We now consider the non-linear regime for the fluctuations in the
presence of an array of ideal SIS junctions. \textit{Below}
$T_c$, the situation of grains coupled through Josephson
junctions was considered in Ref. \cite{clem1987}. It was shown that,
for small grains (as compared to the coherence length), the
classical solution is identical to the one of a dirty
superconductor, provided one uses the effective normal state
resistivity, which incorporates the junction resistance. In this
case, we would expect a conventional behavior in the fluctuation
regime. However, as noticed in Ref. \cite{clem1987}, this
effective medium result should not be valid when thermal or
quantum fluctuations become large enough to destroy phase
coherence between grains. In the present case, the normal state
resistivity at $T_c$ per square and per superconducting plane is high ($R_{\;\Box}\simeq 1.6$ k$\Omega$ for the
optimally doped state, and $R_{\;\Box}\simeq 6.4$ k$\Omega$ for the underdoped state with $T_c = 7.5$ K), as compared to the critical value $R_c=h/4e^2$ = 6.45 k$\Omega$\cite{orr1986}. Then, considering
the additional effect of the charging energy\cite{orr1986b},
there could be in the present case large fluctuations which
contribute to destroy the phase coherence between grains, with a
paracoherent state above $T_c$. As a consequence, the classical
treatment might not be appropriate here.

The non-linear transport properties above $T_c$ is in this case a
largely unexplored field. Kulik derived the expression for the
non-linear excess conductivity of a single tunnel
junction with negligible charging energy\cite{kulik1971}. As
expected, when the resistance of the junction is large
($\epsilon\lesssim R/R_c $), tunneling is dominated by thermal
fluctuations in the junction. The characteristic voltage across
the junction for non linearity is set in this case by the
lifetime of the tunneling pairs in the junction. Further above
$T_c$, the characteristic voltage is set by the pair relaxation
time. In both cases, the temperature sets the energy scale and,
apart from different temperature dependences, the result for the
characteristic electrical field is essentially the same as for the
bulk, $E_c \simeq T_c/e\,\xi$. When coupling is strong enough so
that the correlation length of the array exceeds the separation of
the grains, the array has collective modes\cite{deutscher1973}.
Assuming a uniform flow, the effect of the transport current is a
mere shift of the array free energy, just as a uniform stress on
an harmonic crystal leaves its normal vibrating modes unchanged.
As a consequence, within the Debye
approximation\cite{deutscher1974}, we do not expect that the
current should alter the fluctuations of the array before it breaks the
coupling in the individual junctions.

Then, it seems impossible to explain a substantial reduction of
the characteristic electrical field from a model of SIS junctions
array. So far however, we have not considered the possibility that,
although granular, the junctions are not of the SIS type, but of
the SNS one. In the latter case, \textit{below} $T_c$, the order
parameter induced in the normal metal by the proximity effect is
exponentially reduced with respect to its value in the
superconductor \cite{degennes1964}, as $\exp(-d/\xi_n)$, with $d$
the grain separation and $\xi_n=\hbar\;v_F/2\pi k_B\,T$ the normal
metal-coherence length in the clean limit. Besides reducing the
transition temperature to the decoupling
one\cite{geshkenbein1998}, we expect that, in the fluctuation
regime when $d>\xi$, the characteristic voltage across the
junction should be determined by the effective gap value in the
metal, which is reduced roughly in the same proportion (the
diffusion time in the normal metal being less than the pair
lifetime, $\pi\hbar/8k_B(T-T_c)$). Using $\hbar\;v_F \simeq 1$ eV~\AA, one has $\xi_n\simeq 90 \;\AA$, so that to obtain the measured reduction of
the effective gap value, $\xi\exp(-d/\xi_n)/d$, of two orders of
magnitude, normal domains as large as $d\simeq200$~\AA{} are needed (we have used the clean limit for these estimates, as the mean free path is found $l \gtrsim \xi_n$ from the plasma frequency and the resistivity --- see below). There are, however, objections
to the existence of such normal metal domains. First, such a large
value of the domain size is unlikely, being much larger than the
inhomogeneities which have been put into evidence in Bi$_2$Sr$_2$CaCuO$_{8+\delta}$.
Then, while in the case of Bi-based materials metallic domains
may exist, as overdoping is easy (a possibility not available in
other materials, such as YBa$_2$Cu$_3$O$_x$), we would expect a
reduction of the normal domain size when doping is reduced
(approaching the insulating state), while the effective gap
reduction appears to be uniform with doping.

Several other mechanisms have been proposed that may also
contribute to a decrease of the apparent gap value in junctions
$I-V$ characteristic. Below $T_c$, these mechanisms are invoked to
account for the small $I_c\,R_n$ product generally observed for
grain boundary junctions\cite{hilgenkamp2002}. As pointed out by
Halbritter\cite{halbritter1992}, real junctions in cuprates ---
either natural or artificial --- may be far from the idealized
sharp barriers, which are fully described by the energy barrier
and the tunneling distance. In particular, he has shown that
resonant-states in an insulating barrier may account for a
proximity-like effect, but with a much reduced effective normal
metal coherence length, which can be in the nanometer range.
Identical mechanisms may be at work to weaken the effective
superconducting interaction of the fluctuating tunneling pairs. Such mechanisms are needed to
conciliate the granular description
and the experimental observations.

More generally, it must be noticed that the resonant states mechanism mimics
a \textit{gapless} superconductivity. As stressed in \cite{kaiser1970}, the microscopic
mechanism at the interface is not truly gapless, as both the gap value and the transition
temperature are reduced at the interface, so that the ratio $\Delta/T_c$ is essentially invariant.
However, the granular material made of such junctions does exhibit a reduced $\Delta/T_c$ ratio,
when $\Delta$ is inferred from the nonlinear transport properties. One may wonder if a truly
gapless superconductivity could account for our results. Gapless superconductivity may be
obtained from a bulk scattering mechanism. As well known, \textit{in the case of a d-wave
superconductor}, the superconducting transition temperature, the superfluid density, the
depairing current density, all decrease roughly linearly in a large range of the inverse
scattering rate, $1/\tau$, reaching zero for
$\hbar/2\tau \,k_BT_{c0} \simeq 1$\cite{hirschfeld1993,heesang1994}.
A strong pair-breaking effect is  plausible in the case of Bi-2201.
An estimate for the scattering rate $\tau(T_c)\simeq 0.05$~ps may be obtained,
using $\omega_p=(4\pi\,ne^2/m^*)^{1/2}\simeq 9000$~cm$^{-1}$ as the plasma frequency
\cite{tsvetkov1997} and $\rho\,(T_c)=190\;\mu\Omega$~cm. This yields for the pure material
$T_{c0}\simeq \hbar/2\tau \simeq 10^2$~K and $T_c/T_{c0}\simeq 0.2$. We expect that such a
disordered, gapless superconductor may exhibit a strongly reduced characteristic electrical
field, as this field is determined by the energy spectrum of the excitations, rather than by
the non-zero pair potential\cite{maki1969}. However, scanning tunneling spectroscopy showed
clear evidence for a superconducting gap in Bi-2201, with well defined coherence peaks, but
strongly inhomogeneous at some length scale below 20~nm\cite{kugler2001}. As a consequence,
a bulk, homogeneous description of the disorder in this material, leading to gapless
superconductivity, seems to be inadequate and a granular description more appropriate.
However, in the absence of available theories for the non linear excess conductivity in
disordered d-wave materials, we cannot totally exclude the homogeneous disordered scenario.

\begin{figure}
\resizebox{0.7\columnwidth}{!}{%
\includegraphics{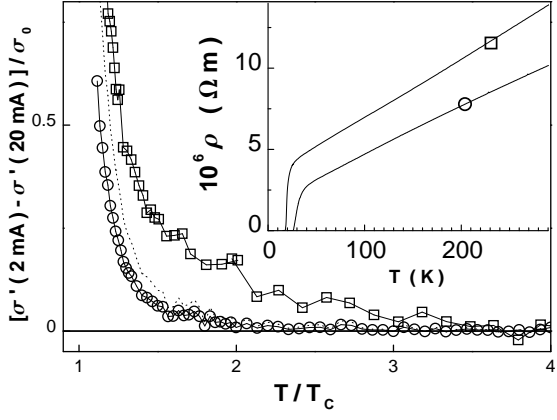}
}
\caption{\label{BiLa}Excess conductivity non linearity. Squares: sample c, optimally doped. Circles: Bi$_2$Sr$_{1.7}$La$_{0.3}$CuO$_{6+x}$, slightly underdoped ($T_c/T_c^{max}=$0.98, where $T_c^{max}$ = 30 K). The dotted line is the corrected data for the La substituted sample by a factor $\rho_{pure}/\rho_{substituted}$, so that both samples may be compared for a constant electrical field.}
\end{figure}

So far, we have considered only the effect of the current on the
fluctuating inter-grain Josephson current, but it is worth
pointing out the possibility of a contribution of the normal state of the granular medium to 
the non linear resistivity. Indeed, there
is an electric field-induced conduction in the case of granular
metals, which has been modeled in
refs.\cite{sheng1972,sheng1973}. The model accounts for both the
exponential decrease of the low temperature conductivity with
electric field, and the activated behavior of the low electric
field regime when transport is dominated by the charging energy
of the grains. However, this picture is difficult to reconcile
with the observation that the non-linearity effect follows the
same doping dependence as the transition temperature and, thus,
is likely related to the superconducting fluctuations rather
than to the normal state. This difficulty may be circumvented
provided one links the occurrence of superconductivity with the
normal state properties themselves. Going to the underdoped
regime, one may expect an increase of the charging energy and of
the barrier height between grains, yielding an increase of the
characteristic electric field \cite{sheng1972,sheng1973},
consistent with our observations. Moreover, assuming that the
superconducting transition is set by the charging energy being of
the order of the Josephson coupling energy, it is found that the
electric field for non linearity is reduced by the WKB tunneling
rate exponential factor, with respect to the intrinsic value\cite{sheng1972,sheng1973}. 
However, within such a picture, it is still difficult to account for the scaling of the non linear effect 
with the critical temperature in the \textit{whole} doping range. 
In a general way, this scaling indicates
that granularity is here likely different from the one observed in
\cite{cren2000,pan2001,howald2001,lang2002}, which was found more
pronounced in the underdoped regime than it is in the overdoped
one. Then, although a similar mechanism should not be excluded in
the case of an intrinsic granularity (such as a real space
electronic phase separation for a slightly doped Mott insulator),
disorder should, in the present case, be attributed to some local
off-stoichiometry or some structural disorder, independent of the
doping level. Also, it is worth to underline that such a disorder apparently brings 
the energy scale for fluctuations suppression to such a low value, that the pseudo-gap phase for this compound \cite{kugler2001} would not be uncovered by the current, even though it could involve superconducting fluctuations.
Concerning the origin of these inhomogeneities, amongst the Bi family, Bi-2201 exhibits a more
pronounced modulation of the BiO plane\cite{zandbergen1988}. Such
a disorder was invoked in Ref. \cite{li2004} to explain the
anomalously low value of the superconducting transition
temperature, $T_c^{max} \simeq 20$ K, where parent compounds such
as Tl$_2$Ba$_2$CuO$_6$ or HgBa$_2$CuO$_4$ have a maximum critical
transition temperature $T_c \simeq$ 90~K. Moreover, when cation
substitution La/Sr is made on Bi-2201, the modulation is reduced
and the maximum transition temperature is increased\cite{li2004}.
The nanoscale domains arising from this modulation could be at
the origin of some granularity in the superconducting plane.
Although there is at present no direct evidence that these
domains induce some inhomogeneity in the superconducting
gap\cite{kugler2001}, it remains that the La substituted material
exhibits a reduced sensitivity to the electrical field. This is
shown in Fig. \ref{BiLa} where the
Bi$_2$Sr$_{1.7}$La$_{0.3}$CuO$_{6+x}$ material, with a residual
resistivity less than the pure material and, presumably, less
disorder, also clearly shows a reduced effect of the electrical
field.

\section{Conclusion}
In conclusion, we have shown that the non linearity with current of the excess-conductivity amplitude for a Bi$_2$Sr$_2$CuO$_{6+\delta}$ thin
film at various doping states, measured well above $T_c$, is
related to the reduced temperature $T/T_c$ only. Its magnitude as well as its temperature dependence are
clearly different from the expectations from the theory of
Gaussian fluctuations in a conventional two-dimensional
superconductor. Based on scanning tunneling measurements, we suggest that this could be the signature of a
granular superconductor, with a gapless-like behavior.

\begin{acknowledgments}

We acknowledge the support of CMCU to project 04/G1307.

\end{acknowledgments}

%
%
%

\newpage

\end{document}